\documentclass{achemso}

\usepackage{amsfonts}
\usepackage{amsmath}
\usepackage{amssymb}
\usepackage{mathbbol}
\usepackage{mathtools}
\usepackage[english]{babel}
\usepackage{float}
\usepackage{graphicx}
\usepackage[hidelinks]{hyperref}
\usepackage{tikz}
\usepackage{adjustbox}
\usepackage{hhline}
\usepackage[T1]{fontenc}

\usepackage{siunitx}
\usepackage[flushleft]{threeparttable}

\definecolor{lime}{HTML}{A6CE39}
\DeclareRobustCommand{\orcidicon}{
  \begin{tikzpicture}
    \draw[lime, fill=lime] (0,0)
    circle [radius=0.14]
    node[white] {{\fontfamily{qag}\selectfont \tiny ID}};
    \draw[white, fill=white] (-0.0625,0.095)
    circle [radius=0.007];
  \end{tikzpicture}
  \hspace{-2mm}
}
\foreach \x in {A, ..., Z}{
  \expandafter\xdef\csname orcid\x\endcsname{\noexpand\href{https://orcid.org/\csname orcidauthor\x\endcsname}{\noexpand\orcidicon}}
}
\def \dehl{\ensuremath{{\Delta E}_\mathrm{HL}}}
\def\desco{\ensuremath{{\Delta E}_\mathrm{sco}}}
\def\thalf{\ensuremath{T_{1/2}}}

\def  \CrII{Cr$^\mathrm{II}$}
\def  \MnII{Mn$^\mathrm{II}$}
\def \MnIII{Mn$^\mathrm{III}$}
\def  \FeII{Fe$^\mathrm{II}$}
\def \FeIII{Fe$^\mathrm{III}$}
\def  \CoII{Co$^\mathrm{II}$}
\def \CoIII{Co$^\mathrm{III}$}

\def \scoA{[Cr(I$_2$)(depe)$_2$]}                      %sco01
\def \scoB{[Mn(L$_1$tren)]}                            %sco02
\def \scoC{[Mn(3,5-diBr-sal$_2$323)](BF$_4$)$\cdot$EtOH} %sco03
\def \scoD{[Mn(L$_2$)](PF$_6$)}                        %sco04
\def \scoE{[Mn(Cp$^{1\mbox{-}\mathrm{Me}}$)$_2$]}      %sco05

\def \scoF{[Mn(Cp$^{1\mbox{-}t\mathrm{Bu}}$)$_2$]}     %sco06
\def \scoG{[Mn(Cp$^{1,3\mbox{-}t\mathrm{Bu}}$)$_2$]}   %sco07
\def \scoH{[Fe(acac)$_2$(trien)](PF$_6$)}              %sco08
\def \scoI{[Fe(qsal-Br)$_2$](NO$_3$)$\cdot$2MeOH}%sco09
\def \scoJ{[Fe(3-OMe-SalEen)$_2$](PF$_6$)}       %sco10

\def \scoK{[Fe(phen)$_2$(NCS)$_2$]}                    %sco11
\def \scoL{[Fe(stpy)$_2$(NCS)$_2$]}                    %sco12
\def \scoM{[Fe(bpp)$_2$](BF$_4$)$_2$}                  %sco13
\def \scoN{[Fe(H$_2$B(pz)$_2$)$_2$(bipy)]}             %sco14
\def \scoO{[Fe(tzpy)$_2$(NCS)$_2$]}                    %sco15

\def \scoP{[Co(terpyridine)$_2$]}                      %sco16
\def \scoQ{[Co(H$_2$(fsa)$_2$en)(pyridine)$_2$]}       %sco17
\def \scoR{[Co(terpyridone)$_2$](ClO$_4$)$_2$}         %sco18
\def \scoS{[Co(MeO-terpy)$_2$](BF$_4$)$_2$}            %sco20
\def \scoT{[Cr(ddpd)$_2$](BF$_4$)$_2$}                 %sco21

\def \scoU{[Cr(octaisopropyl)$_2$]}                    %sco22
\def \scoV{[Co(Tp$^{t\mathrm{Bu},\mathrm{Me}}$)(NAd)]} %sco23
\def \scoW{[Co($^\mathrm{Ar}$L)(N$^t$Bu)]}             %sco24
\def \scoX{[Fe(\emph{t}Bu$_2$qsal)$_2$]}               %sco25

\def \extD{[Mn(L$_2$)]$^{+}$}                          %ext04
\def \extH{[Fe(acac)$_2$(trien)]$^{+}$}                %ext08
\def \extM{[Fe(bpp)$_2$]$^{2+}$}                       %ext13
\def \extT{[Fe(H$_2$O)$_6$]$^{3+}$}                    %ext21
\def \extU{[Fe(eda)$_3$]$^{3+}$}                       %ext22
\def \extV{[Fe(tacn)$_2$]$^{2+}$}                      %ext23

\title{How Accurately Can We Describe Spin Crossover?}

\author{Angel Albavera-Mata\orcidA{}}
\affiliation{Department of Physics,
  University of Florida, Gainesville, Florida 32611}
\alsoaffiliation{Center for Molecular Magnetic Quantum Materials, Gainesville, Florida 32611}

\author{Daniel Mej{\'i}a-Rodr{\'i}guez\orcidB{}}
\affiliation{Physical Sciences Division,
  Pacific Northwest National Laboratory, Richland, Washington 99354, United States}

\author{Niranjan Govind\orcidC{}}
\affiliation{Physical Sciences Division,
  Pacific Northwest National Laboratory, Richland, Washington 99354, United States}
\alsoaffiliation{Department of Chemistry,
  University of Washington, Seattle, Washington 98195}

\author{Ajay Panyala\orcidD{}}
\affiliation{Advanced Computing, Mathematics, and Data Division,
  Pacific Northwest National Laboratory, Richland, Washington 99354, United States}

\author{Richard G. Hennig\orcidE{}}
\email{rhennig@ufl.edu}
\affiliation{Department of Materials Science and Engineering,
  University of Florida, Gainesville, Florida 32611}
\alsoaffiliation{Center for Molecular Magnetic Quantum Materials, Gainesville, Florida 32611}

\author{S.B. Trickey\orcidF{}}
\email{trickey@ufl.edu}
\affiliation{Department of Physics and Department of Chemistry,
  University of Florida, Gainesville, Florida 32611}
\alsoaffiliation{Center for Molecular Magnetic Quantum Materials, Gainesville, Florida 32611}

\abbreviations{GGA, generalized gradient density functional approximation;
               CCSD, coupled cluster with single and double excitations;
               CCSD(T), coupled cluster with single, double, and perturbative triple excitations}


\begin{document}

\newpage

%%%%%%%%%%%%%%%%%%%%%%%%%%%%%%%%%%%%%%%%%%%%%%%%%%%%%%%%%%%%%%%%%%%%%%%%%%%%%%%%

\begin{tocentry}
  \includegraphics[width=\columnwidth]{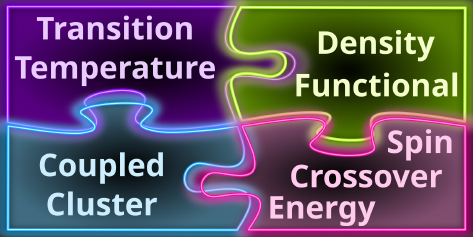}
\end{tocentry}

%%%%%%%%%%%%%%%%%%%%%%%%%%%%%%%%%%%%%%%%%%%%%%%%%%%%%%%%%%%%%%%%%%%%%%%%%%%%%%%%

\begin{abstract}
  The complicated physicochemical properties of metal complexes that exhibit
thermal spin crossover make it difficult for routine electronic structure
calculations to yield an accurate transition temperature prediction,
{\thalf}. The difficulty lies in the intricate connection
between the spin-crossover energy, which is a
molecular spectroscopic property, and {\thalf}, a condensed phase
property. Here we show how to obtain spin-crossover energies systematically by reverse
engineering of experimental {\thalf} data. The protocol is based upon fitting the
range separation parameter, $\omega$, in the hybrid LC-$\omega$PBE
density functional to reproduce the experimental {\thalf} values for a
series of metal complexes. We provide insights into the sources of
variations of at least $\pm 15$ kJ\,mol$^{-1}$ found from common exchange
and correlation functionals by comparing their performance against our
reference data. By analysis of the sensitivity of transition temperatures
to $\pm 1$ \% shifts in the range separation parameter, we determined a
typical uncertainty of $\pm 50$ K for them, and a $\pm 2$ kJ\,mol$^{-1}$
uncertainty in the extracted spin-crossover energies due to $\pm 1$ \%
variations of {\thalf}. Lastly, we present results from the high-level,
all-electron coupled cluster method for eight of the smaller molecules in
the reference data set, and discuss the influence of the truncation of
the excitation series upon the spin state energies.
\end{abstract}

%%%%%%%%%%%%%%%%%%%%%%%%%%%%%%%%%%%%%%%%%%%%%%%%%%%%%%%%%%%%%%%%%%%%%%%%%%%%%%%%
\section{Introduction \label{sec:introduction}}
%%%%%%%%%%%%%%%%%%%%%%%%%%%%%%%%%%%%%%%%%%%%%%%%%%%%%%%%%%%%%%%%%%%%%%%%%%%%%%%%

  The spin crossover phenomenon long has been known and studied intensively by
experimenters.\cite{P1931, CS1931, BB1964, KM1966} It is observed primarily
in condensed phase samples of first transition row $3d^4$ to $3d^7$ metal
complexes. Those switch from a low- to a high-spin state in the presence of
some external stimulus, most often temperature, though pressure and light also
are prominent in the literature.\cite{K1993, GG2004, GG2004I, GG2004II,LGG2004,
J2005, M2006, DR2016} That spin-state switching typically is accompanied by
changes in the electric, magnetic, or optical properties that make these
molecules attractive for use as sensors,\cite{GKG2002, JLRRPDWG2015} memory
storage,\cite{KKJ1992, KM1998, LSGTMMB2008} and actuators,\cite{MRSMQNSB2016,
MMSRRSBCMCLNNSMB2018, YTMSB2026} among other applications.\cite{MVL2003, S2011,
QFSSMSNB2014, R2014, CP2021, NCRP2022,ASSAIHTH2023}

The molecular mechanism that underlies this condensed phase spin-state
conversion involves a delicate competition between the spin pairing energy,
$\mathcal{P}$, and the ligand field splitting, $\Delta$, of the $d$ orbitals due
to the molecular symmetry and ligand strength. That competition ultimately
dictates whether the occupation of the $d$-manifold favors maximum spin
multiplicity according to Hund’s rule. For strong ligand fields, $\Delta >
{\mathcal P}$ and electrons preferentially occupy the lower-energy orbitals,
leading to a low-spin state. Conversely, with weak ligand fields, ${\mathcal P} >
\Delta$, the electrons occupy the maximum number of orbitals according to Hund's
rule, resulting in a high-spin state. Crucially, when $\Delta \approx \mathcal{P}$,
a small external perturbation can overcome either effect and induce reversible
spin-state switching.\cite{GG2004I, B2015, RMSNB2017, NB2018, HH2026}

  For thermally driven transitions in particular, the Gibbs free energy $G =
H - T S$ describes the spin conversion. The enthalpic contribution, $H$, commonly
is related to the distortions that arise from the increase in the metal--ligand
bond length upon switching from a low- to a high-spin state and vice versa,
whereas the entropic contribution, $S$, arises from the extent of both structural
disorder and electronic population characteristic of the high-spin state, and,
importantly, the differences in vibrational spectra between the two spin
states.\cite{NB2018, HH2026} 

  In the experimental  context,\cite{P2021} it is possible to find
$N_\mathrm{HS}$ high-spin state molecules at temperature $T$ among a total of
$N$ weakly interacting species in a bulk molecular crystal. It follows that
the Gibbs free energy of an ideal solution model is expressed in terms of the
Gibbs free energy of the individual spin states, $G_\mathrm{LS}$ and
$G_\mathrm{HS}$, weighted by the relative high-spin state fraction
$n_\mathrm{HS} = N_\mathrm{HS}/N$, along with the ideal entropy of mixing
$S_\mathrm{mix} = -k_B\,N_A\,(\; n_\mathrm{HS} \ln[n_\mathrm{HS}] +
(1 - n_\mathrm{HS}) \ln[1 - n_\mathrm{HS}]\;)$ in the form
\begin{equation} \label{eq:1}
  G = n_\mathrm{HS}\,G_\mathrm{HS} +
      (1 - n_\mathrm{HS})\,G_\mathrm{LS} -
      \, T S_\mathrm{mix}
\end{equation}
The equilibrium condition ${\left(\partial G/\partial n_\mathrm{HS}\right)}_{T,P} = 0$
defines the transition temperature
\begin{equation} \label{eq:2}
  \thalf = \frac{\Delta H}{\Delta S}
\end{equation}
that corresponds to the extremum $n_\mathrm{HS} = 1/2$, namely, an equal
population of both low- and high-spin states. We alert the reader that eq
\eqref{eq:2} is valid at $n_\mathrm{HS} = 1/2$ where the derivative of the ideal
mixing entropy with respect to $n_\mathrm{HS}$ vanishes, allowing $\Delta H$ and
$\Delta S$ to be interpreted as molecular thermodynamic quantities. As a result,
the enthalpic term $\Delta H$ that stabilizes the low-spin state is dominant for
$T < \thalf$, whereas the entropic term $\Delta S$ that stabilizes the high-spin
state prevails for $T > \thalf$.\cite{GG2004, NB2018}

 The challenge for theory and computation is in
$\Delta H = \desco + \Delta E_\mathrm{vib} + P\,\Delta V$, with $P$ the
pressure, $V$ the volume, $E_\mathrm{vib}$ the molecular and condensed phase
vibrational energy, and {\desco} the molecular spin-crossover energy.
It is difficult to extract {\desco} experimentally, so
reported experimental values are rare and depend upon diverse protocols.
What we construct and use here is a systematic alternative,
namely a well-defined theory and computation protocol for extracting {\desco}
from experimental {\thalf} data.

  For context, both theory and detailed computational
studies have been done.\cite{KL2010, PSW2013, RHBM2020, R2023, RVK2024,
GV2026} Those include thermodynamic models\cite{SD1972, BL2003, H2013, RMSNB2017}
and detailed microscopic models\cite{PB2013, PL2018, OPDLK2018, EN2018} rooted
in some form of electronic structure methodology. In them, the
largest contribution to the Gibbs free energy comes from molecular total
electronic energies. Specifically, the spin gap energy $\dehl = E_\mathrm{HS} -
E_\mathrm{LS}$ is the total energy difference between the high- and
low-spin states, $E_\mathrm{HS}$ and $E_\mathrm{LS}$, respectively. As explained
in ref \citenum{A2025}, inclusion of the zero-point energy
difference, ${\Delta E}_\mathrm{zpe}$, results in the molecular
spin-crossover energy {\desco} given as
\begin{equation} \label{eq:3}
  \desco = \dehl + {\Delta E}_\mathrm{zpe}
\end{equation}

  Accurate electronic structure determination of the {\desco} magnitude and sign
is key for trustworthy prediction of {\thalf}.\cite{PSW2013,RHBM2020} While not
obligatory, Kohn-Sham calculations with various exchange and correlation
functional approximations are dominant. From here onward we denote them
simply as functionals. Because of the delicate balance mentioned above, hybrid
functionals \cite{K2016, CVR2018, VCR2021} often are used.
Promising results have been reported and might be 
considered as starting choices for large-scale, high-throughput calculations.
In addition, however, to issues of computational cost, the choice of functional often 
can be a matter of computational art. To enable interpretive accuracy, the
physicochemical nature of each metal complex typically is taken into account
first by choosing a functional with presumably desirable properties, then either by
tuning a mixing parameter for the proportion of single determinant exchange if
the choice is a hybrid, or by choosing the magnitude of an on-site Coulomb
interaction in the density functional theory plus Hubbard-$U$
method.\cite{VFRR2015, VFRR2020, OP2020, AHT2022} Perhaps unsurprisingly, in the
comparatively few cases that an experimental estimate of {\desco} is given, the
situation is essentially an exact parallel. As noted already, the analysis
extracting that {\desco} value from the measured {\thalf} is specific to the
particular experimental technique.

  An alternative electronic structure methodology that makes no use of
density functional approximations is the
coupled cluster method.\cite{C1958, CK1960, C1966, PCS1972, BP1978, PB1982,
RTPH1989, PL1999} In practice, it constructs a hierarchy of many-electron wave
functions based on increasing numbers of excitations, namely, single, double,
triple, etc., with respect to a single-determinant reference state. Because
the method has a costly scaling with the number of electrons, it is normal
practice to do full single and double excitations, denoted CCSD, and add
perturbative triple excitations, CCSD(T). This is considered the gold standard
for molecular electronic structure calculations. For large metal complexes in
particular, the costly scaling with particle number poses a serious limitation.
Surrogate, lower-scaling coupled-cluster approximations, such as the domain-based
local pair natural orbital method,\cite{RN2013} have been applied to such large
systems,\cite{NPM2020, FGRTN2020, CFKMR2020, MLH2020, CNV2023} but with several
works showing sensitivity to the cutoff choices.\cite{DMP2022, RK2022, PSK2023,
ARNB2023}

  Here we explore both approaches from what can be viewed as an inverse
data-driven perspective.\cite{R2023} The strategy is to generate a reference {\desco} data
set that is derived from experimental {\thalf} values via a systematic, reproducible
theory-computation protocol. With that data set as the reference, then we
investigate methodological  sensitivity to procedural and computational choices,
compare the performance of different density functionals, and test the utility
of straight-forward but exascale CCSD(T) calculations.

  Our theory-computation protocol is similar in concept to previous
schemes that make use of experimental properties for tuning parameters in
functionals.\cite{SKB2009, BLS2010, SEKB2010, SJCDSS2025} Those works
involve a different tuning, namely to satisfy the density functional theory
ionization potential theorem. Ref \citenum{SYMBY2014}, however, investigated
a scheme whose basis is similar conceptually to ours. Their goal was to 
calibrate certain functionals to support detailed computational
investigation of specific complexes. Hence, they re-parametrized the
ordinary linearly mixed hybrid functionals B3LYP\cite{B1993,SDCF1994} and
PBE0\cite{ES1999, AB1999} by tuning the mixing parameter to yield a
match of the calculated and experimental {\thalf} values. Their example systems
were two Fe(II) spin-crossover complexes. That calibration approach does not
take advantage of the opportunity to tune a rigorously justified hybrid
functional. In that regard, note that PBE0 is a fixed mixture, while B3LYP
is a combination of exchange parmetrization for a different functional and a
correlation approximation to what is known to be a non-$N$-representable
second-order reduced density matrix.\cite{M1993}

  Our approach starts by selecting a long-range corrected hybrid functional.
Such functionals have a range separation parameter that influences the electronic
energy levels directly. The separation is rigorous and unambiguous. We
tune that hybridization parameter to reproduce the experimental {\thalf}
for each spin-crossover metal complex in a set of twenty four. The electronic
structure calculations for that tuning provide a reverse engineering, i.e.
data inversion, means to extract the reference {\desco} associated with the
individual {\thalf}. Via small variations of the hybridization parameter
around its optimal value, we determine rational error bars. With that reference
{\desco} data set in hand, then we test the spin-crossover energies
calculated from several other functionals of distinct refinement and
complexity against the reference data. We find that those functionals
typically show large deviations from the reference values, over $\pm 15$
kJ\,mol$^{-1}$, or nearly 100 \% relative error.

  As a matter of quantitative context, CCSD(T) calculations also are relevant. 
Even with access to forefront exascale machines, computational demands limit us to
treating eight metal complexes with  comparatively small electron counts. Our
results from all-electron CCSD(T) calculations demonstrate effects traceable to the
conventional Hartree-Fock reference state and to the truncation to perturbative
triple excitations. Coupled-cluster treatment of spin-crossover thus is shown to
require a non-conventional reference state or perhaps a multi-reference treatment
and may require higher-order terms for acceptable accuracy as well.

%%%%%%%%%%%%%%%%%%%%%%%%%%%%%%%%%%%%%%%%%%%%%%%%%%%%%%%%%%%%%%%%%%%%%%%%%%%%%%%%
\section{Methods \label{sec:methods}}
%%%%%%%%%%%%%%%%%%%%%%%%%%%%%%%%%%%%%%%%%%%%%%%%%%%%%%%%%%%%%%%%%%%%%%%%%%%%%%%%

  We assembled a data set of twenty-four metal complexes from ref
\citenum{AHT2022} that are known experimentally for exhibiting a thermally driven
spin-crossover transition. For each of those, we performed electronic structure
calculations with the ORCA software package\cite{N2012} using the def2-QZVP basis
sets for Cr, Mn, Fe, and Co, and the def2-TZVP basis sets otherwise.\cite{WA2005}
The integration quadratures used Becke weights and angular Lebedev grid schemes
with 110, 434, 590, 770, and 590 points for the five-region partitioning in ORCA,
respectively, with adaptive pruning. In addition, we chose a general integration
accuracy of six to determine the number of radial points, and a basis function
cutoff of $10^{-12}$ for the numerical integration. At this point it is
important to highlight that we used variational Coulomb fitting,\cite{BID2010}
also known as resolution of the identity,\cite{N2003, W2006} to calculate
Coulomb repulsion integrals, and the chain-of-spheres approximation for the
single determinant exchange in hybrid density functional approximations.\cite{W2007,
HSNI2021} To avoid converging to a spurious electronic excited state, level
shifting of the unoccupied diagonal elements of the Fock matrix was completely
deactivated. Various threshold choices included setting the self-consistent
field energy convergence to $10^{-9}$ atomic units, while the root mean square and
maximum change for the electron density were fixed to $10^{-8}$ and $10^{-9}$ atomic units,
respectively. The convergence tolerance for the direct inversion in the
iterative subspace was set to $10^{-8}$, with primitive integral prescreening
cutoff of $10^{-14}$. Finally, the orbital gradient and orbital rotation angle
convergences both were set to $10^{-6}$.

  The reference data was computed with the LC-$\omega$PBE\cite{VS2006, VHKS2006}
range-separated generalized gradient density functional approximation, and the
D3 dispersion correction method using the Becke-Johnson damping
function.\cite{GEG2011, GG2011} Motivations for this choice are discussed
in the next section. We fit the $\omega$ parameter for each metal
complex to reproduce as closely as possible the reported experimental transition
temperature {\thalf}. For that purpose, we screened a series of choices for $\omega$. A
plausible initial value was taken, followed by variations in steps of
10$^{-2}$, each involving a geometry relaxation and a frequency analysis for
both the low- and high-spin states to determine the associated {\thalf}. After
that we fit {\thalf} as a function of $\omega$ to determine the $\omega$ value
that reproduced, as well as possible, the experimental transition temperature.
Then we ran a final geometry relaxation, followed by a harmonic frequency analysis
for both spin states using that tuned $\omega$. This procedure was used for
each species in the data set. In each case we confirmed the absence of imaginary
harmonic modes. The geometry relaxations were carried out with the limited-memory
Broyden, Fletcher, Goldfarb, and Shanno scheme.\cite{LN1989} We used a threshold
of $10^{-5}$ for both the maximum gradient and maximum displacement, and
$10^{-4}$ for the root mean squared of both the gradient and the displacement.
Regarding the harmonic frequency analyses, we used a tolerance of $10^{-10}$ for
the coupled perturbed self consistent field with the Pople solver.

  A procedural nuance regarding the calculation of {\thalf} is worth discussing
at this point. As remarked in the Introduction, the experimental transition
temperature, as measured, is a condensed phase datum, whereas our molecular
approach accounts only for the vibrational contributions associated with molecular
degrees of freedom. The molecular approach results in mean {\dehl}
differences of roughly 4.3 and 2.4 kJ\,mol$^{-1}$ for PBE and r$^2$SCAN,
respectively, with respect to the condensed phase data.\cite{AHT2023}
Isolated molecules cannot account for phonon contributions,
packing effects, inter-molecular interactions, nor spin-lattice coupling.
Several studies have attempted to include solid-state effects through
periodic Kohn-Sham calculations,\cite{PSW2013, H2013} elastic continuum
models,\cite{PB2013, KL2010} and explicit spin-lattice coupling
approaches.\cite{PB2013, BL2003} While such treatments are valuable, they
introduce additional computational complexity and parameters that are beyond
the scope of the present molecular-focused benchmarking study. Indeed, one
can make the case that because these metal complexes occur in
molecular crystals, the intra-molecular motions may be considered as the dominant
degrees of freedom, with the lattice modes being much less important in
contributions to equation \eqref{eq:2}.

  Moving on, exchange and correlation functionals were chosen for comparison
purposes using single-point energy calculations, each with its D3 dispersion
correction parameters for the Becke-Johnson damping function.\cite{GEG2011,
GG2011, KM2013, LLMC2013, MG2014, BBSP2016, GHBENG2017, EHNFSKPB2021} Those
single point tests used the computational choices described above and the
geometries that resulted from fitting the $\omega$ parameter in LC-$\omega$PBE.
The functionals scrutinized were the generalized gradient approximations PBE,\cite{PBE1996,
PBE1996e} B88P86,\cite{B1988, P1986, P1986e, TSH1999} OPBE,\cite{PBE1996, HC2001}
OLYP,\cite{LYP1988, MSSP1989, HC2001} and BLYP;\cite{B1988, LYP1988, MSSP1989}
the meta-generalized gradient approximations TPSS,\cite{TPSS2003}
revTPSS,\cite{PRCCS2009, PRCCS2011}, M06-L,\cite{ZT2008}\footnote{We used the
D3(0) scheme\cite{GEG2011} instead for this density functional as recommended in
ref \citenum{HBGPWD2014}.} SCAN,\cite{SRP2015} rSCAN,\cite{BY2019} and
r$^2$SCAN;\cite{FKNPS2020} the global hybrid generalized gradient approximations
PBE0\cite{ES1999, AB1999}, B3LYP\cite{B1993, SDCF1994},
and B3LYP$\ast$\cite{RSA2001}; the global hybrid
meta-generalized gradient approximations TPSSh,\cite{SSTP2003}
revTPSSh,\cite{CPR2010} and PW6B95;\cite{ZT2005} the range-separated hybrids
HSE06,\cite{HSE2003, HSE2006, KVIS2006} CAM-B3LYP,\cite{YTH2004} and
$\omega$B97X-D;\cite{CH2008} and the double hybrids
PWPB95\cite{GG2011a} and DSD-PBEB95.\cite{KM2013} Also we included the
PBE+$U_\mathit{eff}$ combination to assess the performance of the
mean-value ensemble method\cite{AHT2022} for the calculation of
effective Hubbard-$U$ values. For those calculations, we kept the
effective $U$ values and computational choices reported in
ref \citenum{AHT2022}, although we included the D3 dispersion
correction and used the reference geometries in this work instead. The
rationale is that the structural differences account only
for small alterations on the order of meV of the effective Hubbard-$U$
values, hence the relevant calculated energy differences will not be
affected significantly. We clarify further that the SCAN and rSCAN functionals
are included solely as complement to r$^2$SCAN. These therefore are not part of
the main discussion of comparative performance.

  Calculations with each of those functionals generated ORCA output files that
then were processed with the pySCO library.\cite{A2025} That included the
computation of the thermodynamic parameters {\thalf}, $\Delta H$, and $\Delta S$,
as well as the energy differences {\dehl} and {\desco}.

  As mentioned already, computational costs limited the all-electron
coupled-cluster calculations to eight relatively small metal complexes, namely,
{\extD},\cite{MGHLNMKM2012}, {\scoE},\cite{SWRM1974} {\scoF},\cite{WSBA2009}
{\extH},\cite{SSDTW1978} {\extM},\cite{CKLH2007} {\extT},\cite{R2013, RGSB2016,
RD2018, FCKV2022}, {\extU},\cite{RB1968, C1994} and {\extV}.\cite{BFHW1985} We
used the ExaChem code,\cite{Exachem2021, Exachem2023} with the cc-pVQZ basis
sets for the transition metals and cc-pVTZ for all others.\cite{D1989, WD1993,
BP2005} The coupled-cluster calculations were single point, with geometries
obtained from structural relaxations with the meta-generalized r$^2$SCAN
functional for both the low- and high-spin states for each metal complex. The
coupled-cluster reference state was unrestricted Hartree-Fock with orbitals
computed using a convergence threshold of $10^{-7}$ atomic units for the total
energy, and $10^{-6}$ for the density, and a tolerance of $10^{-12}$ for the
Schwarz screening for the Coulomb integrals. In addition, we selected a threshold
of $10^{-22}$ for integral primitive screening, while using a threshold of
$10^{-8}$ for the detection of linear dependencies given by our choices of basis
sets. Lastly, we used $10^{-6}$ as the convergence criterion for the residual
norm of the amplitude equations for the post-Hartree-Fock calculations, and we
included a level shifting of $10^{-1}$ atomic units in an effort to stabilize
the convergence of the cluster amplitudes.

%%%%%%%%%%%%%%%%%%%%%%%%%%%%%%%%%%%%%%%%%%%%%%%%%%%%%%%%%%%%%%%%%%%%%%%%%%%%%%%%
\section{Discussion \label{sec:discussion}}
%%%%%%%%%%%%%%%%%%%%%%%%%%%%%%%%%%%%%%%%%%%%%%%%%%%%%%%%%%%%%%%%%%%%%%%%%%%%%%%%

  As remarked in the Introduction, in the context of comparatively sophisticated
functionals, the unique physicochemical character of each metal complex means
that reproducing experimental results requires either a specific contribution of
single determinant exchange or range separation tailored to the species at hand.
On that account, we chose the LC-$\omega$PBE range-separated hybrid. The choice
is motivated by the fact that LC-$\omega$PBE is based on the widely used PBE
functional, yet involves only one adjustable parameter, $\omega$. It controls
the separation regime in a physically and formally motivated way that is
distinct from the fixed-ratio separation in global hybrids.
That in turn is connected directly to the spin manifold
energetics. This is a major advantage for our purposes, since we may control the
$\omega$ magnitude so as to replicate an experimental property of interest, then
analyze the corresponding spin manifold.

\begin{table}
\centering
  \caption{Metal-organic complexes in the data set sorted by metallic ion. The
           $\omega$ parameter, in units of Bohr$^{-1}$, in the LC-$\omega$PBE
           range-separated hybrid was fitted for each metal complex to reproduce
           as closely as possible the experimental transition temperature
           $\thalf^\mathrm{ref}$. The spin gap energy, and the spin-crossover
           energy that includes the zero-point energy correction, $\dehl =
           E_\mathrm{HS} - E_\mathrm{LS}$ and $\desco = \dehl +
           \Delta E_\mathrm{zpe}$, respectively, are shown in units of
           kJ\,mol$^{-1}$. The reference $\thalf^\mathrm{ref}$ and calculated
           transition temperatures $T_{1/2}$, and the associated changes in
           enthalpy, $\Delta H$, and entropy, $\Delta S$, are shown in units of
           Kelvin, kJ\,mol$^{-1}$, and J\,mol$^{-1}$\,K$^{-1}$, respectively.
           \label{tab:I}}
  \begin{adjustbox}{max width=\columnwidth}
  \begin{threeparttable}
  \begin{tabular}{lcc
                  S[table-format=3.1]l
                  S[table-format=1.4]
                  S[table-format=2.2]
                  S[table-format=2.2]l
                  S[table-format=3.0]
                  S[table-format=2.2]
                  S[table-format=2.2]}
    \hline
    \multicolumn{1}{c}{Complex}&Ion&Ref&{$\thalf^\mathrm{ref}$}&&{$\omega$}&{\dehl}&{\desco}&&{\thalf}&{$\Delta H$}&{$\Delta S$}\\ \cline{1-4}\cline{6-8}\cline{10-12}
    {\scoA}&{\CrII} &\citenum{HHLLPS1989}  &171&&0.3437&15.11& 9.42&&171.6&12.94&75.40\\
    {\scoT}&{\CrII} &\citenum{BFCBHHGR2020}&293&&0.1774& 6.97& 5.24&&293.0& 7.11&24.26\\
    {\scoU}&{\CrII} &\citenum{SSDK1997}    &300&&0.4141&18.54&10.20&&292.6&13.54&46.26\\
    {\scoB}&{\MnIII}&\citenum{SS1981}      & 45&&0.4440& 5.91& 0.64&& 38.6& 0.79&20.56\\
    {\scoC}&{\MnIII}&\citenum{PGMHMM2012}  &175&&0.4693&11.60& 2.91&&179.8& 4.64&25.82\\
    {\scoD}&{\MnIII}&\citenum{MGHLNMKM2012}&131&&0.3973&11.53& 3.48&&129.8& 4.89&37.66\\
    {\scoE}&{\MnII} &\citenum{SWRM1974}    &303&&0.4150&24.69&14.89&&308.5&19.32&62.63\\
    {\scoF}&{\MnII} &\citenum{WSBA2009}    &215&&0.4248&22.39&12.25&&216.4&15.62&72.18\\
    {\scoG}&{\MnII} &\citenum{WSBA2009}    &300&&0.4085&30.08&19.04&&288.8&23.35&80.85\\
    {\scoH}&{\FeIII}&\citenum{SSDTW1978}   &200&&0.4589&14.73& 6.70&&208.2& 8.62&41.42\\
    {\scoI}&{\FeIII}&\citenum{HPHMMJ2015}  &136&&0.4738&11.01& 3.39&&139.7& 4.58&32.77\\
    {\scoJ}&{\FeIII}&\citenum{TBCTB2011}   &164&&0.5275&19.07& 8.34&&164.3&10.73&65.26\\
    {\scoK}&{\FeII} &\citenum{GRHZ1990}    &177&&0.2847&15.07& 4.79&&178.9& 7.91&44.22\\
    {\scoL}&{\FeII} &\citenum{RZGGC1994}   &109&&0.3074&20.02& 6.93&&108.7& 9.72&89.42\\
    {\scoM}&{\FeII} &\citenum{CKLH2007}    &256&&0.3187&20.59&12.35&&255.0&16.33&64.06\\
    {\scoN}&{\FeII} &\citenum{RMFS1997}    &160&&0.3386&22.96&10.62&&158.3&14.49&91.53\\
    {\scoO}&{\FeII} &\citenum{NGMABR2003}  &118&&0.2595&12.83& 3.00&&116.0& 4.78&41.18\\
    {\scoX}&{\FeII} &\citenum{GJLUDKJJ2021}&123&&0.3015&12.76& 3.58&&139.4& 5.36&38.43\\
    {\scoV}&{\CoIII}&\citenum{SYZRT2005}   &280&&0.2034&18.28&11.89&&280.7&14.04&50.01\\
    {\scoW}&{\CoIII}&\citenum{KSB2012}     &300&&0.2784& 9.30& 6.72&&308.2& 7.33&23.78\\
    {\scoP}&{\CoII} &\citenum{FKW1983}     &200&&0.3740& 9.61& 5.40&&198.8& 6.98&35.10\\
    {\scoQ}&{\CoII} &\citenum{CNVLTZD1988} &121&&0.3462& 4.79& 1.16&&146.0& 1.43& 9.79\\
    {\scoR}&{\CoII} &\citenum{GMNR2001}    &172&&0.2757& 5.84& 2.92&&188.2& 3.35&17.82\\
    {\scoS}&{\CoII} &\citenum{HNOANOYK2015}&250&&0.3675& 7.25& 3.92&&258.9& 5.37&20.74\\
    \hline
  \end{tabular}
  \begin{tablenotes}
  \setlength\labelsep{0pt}
  \item\scriptsize{depe = 1,2\--bis\-(di\-ethyl\-phos\-phi\-no)e\-tha\-ne,
    ddpd = N,N$^\prime$\--di\-me\-thyl-N,N$^\prime$-di\-pyr\-i\-di\-ne-2-yl-pyr\-i\-di\-ne-2,6-di\-a\-mi\-ne,
    L$_1$tren = tris\-(2\--((py\-rrol-2yl)\-me\-thyl\-ene\-a\-mi\-no)e\-thyl)\-a\-mi\-ne,
    3,5\--di\-Br\--sal$_2$323 = 2,2$^\prime$\--(2,6,9,13\--te\-tra\-aza\-te\-tra\-de\-ca-1,13-di\-ene-1,14\--di\-yl)\-bis\-(4,6\--di\-bro\-mo\-phe\-no\-la\-to),
    L$_2$ = (2,2$^\prime$\--(2,6,9,13\--te\-tra\-aza\-te\-tra\-de\-ca\--1,13\--di\-ene\--1,14\--di\-yl)\-di\-phe\-nol),
    Cp$^{1\mbox{-}\mathrm{Me}}$ = me\-thyl\-cy\-clo\-pen\-ta\-di\-ene,
    Cp$^{1\mbox{-}t\mathrm{Bu}}$ = \emph{tert}\--bu\-tyl\-cy\-clo\-pen\-ta\-di\-enyl,
    Cp$^{1,3\mbox{-}t\mathrm{Bu}}$ = \emph{tert}\--bu\-tyl\-cy\-clo\-pen\-ta\--1,3\--di\-ene,
    acac = a\-ce\-tyl\-a\-ce\-to\-na\-te, tr\-ien = tri\-e\-thyl\-ene\-te\-tra\-mi\-ne,
    qsal\--Br = (N\--8\--qui\-no\--lyl)\--5\--Br\--sa\-li\-cyl\-al\-di\-mi\-na\-te,
    3\--OMe\--SalEen = (2\--(((2\--(e\-thyl\-a\-mi\-no)\-e\-thyl)\-i\-mi\-no)\-me\-thyl)\--6\--me\-tho\-xy\-phe\-no\-la\-to-N,N$^\prime$,O),
    phen = phe\-nan\-thro\-li\-ne,
    stpy = 4\--sty\-ryl\-pyr\-i\-di\-ne,
    bpp = (2,6\--di(py\-ra\-zol\--1\--yl)\-py\-ri\-di\-ne),
    H$_2$B(pz)$_2$ = di\-hy\-dro\-gen bis\-(py\-ra\-zol\--1\--yl)\-bo\-ra\-te,
    bipy = bi\-py\-ri\-di\-ne, tzpy = (3\--(2\--py\-ri\-dyl)\-(1,2,3)\-tri\-a\-zo\-lo(1,5\--a)\-py\-ri\-di\-ne),
    \emph{t}Bu$_2$qsal = 2,4\--di(\emph{tert}\--bu\-tyl)\--6\--((qui\-no\-li\-ne-8-y\-li\-mi\-no)\-me\-thyl)\-phe\-nol,
    Tp$^{t\mathrm{Bu},\mathrm{Me}}$ = hy\-dro\-tris(3\--\emph{tert}\--bu\-tyl,5-methyl$^\prime$\--pyr\-a\-zol\-yl)bo\-ra\-te,
    NAd = a\-da\-man\-tyl az\-ide,
    $^\mathrm{Ar}$L = 5\--me\-si\-tyl-1,9-(2,4,6\--tri\-phe\-nyl\-phen\-yl)di\-pyrro\-me\-the\-ne,
    H$_2$(fsa)$_2$en = 3\--for\-myl\-sa\-li\-cy\-lic a\-cid\--e\-thyl\-en\-di\-a\-mi\-ne,
    MeO\--ter\-py = 4$^\prime$\--me\-tho\-xy\--2,2$^\prime$:6$^\prime$\--ter\-pyr\-i\-di\-ne.}
  \end{tablenotes}
  \end{threeparttable}
  \end{adjustbox}
\end{table}
  We therefore begin the discussion  with the $\omega$ values for the
LC-$\omega$PBE functional tuned to reproduce the experimental {\thalf}. All
the data needed to reproduce our results is included as part of the Electronic
Supplementary Information. Table \ref{tab:I} shows that the fitted $\omega$
magnitudes span roughly the interval $0.177 \leq \omega \leq 0.528$. Values for
most of the systems are centered around 0.36, with the {\CrII} {\scoT} and the
{\FeIII} {\scoJ} complexes having the minimum and maximum values, respectively.
The diverse chemical space of the ligands associated with the metal complexes and
the somewhat limited number of complexes makes it difficult to pick out any
clear correlation between a molecule and its fitted $\omega$. The $\omega$
magnitudes seem independent of the metallic ion as well.

  There are, however, the {\MnII} manganocene analogues {\scoE}, {\scoF}, and
{\scoG} that share similar $\omega$ values, namely, 0.415, 0.425, and 0.409,
respectively. The three ligands with progressively increasing size influence the
final $\omega$. For instance, the \emph{tert}\--bu\-tyl group, \emph{t}Bu, in
the cy\-clo\-pen\-ta\-di\-enyl rings resulted in an $\omega$ slightly larger
than that with the methyl group, Me. Addition of two such \emph{t}Bu substituents
in the cy\-clo\-pen\-ta\--1,3\--di\-ene ring instead resulted in an $\omega$
slightly smaller than that for the same Me group. Notice that the trend for
these $\omega$ values is Cp$^{1,3\mbox{-}t\mathrm{Bu}}$ <
Cp$^{1\mbox{-}\mathrm{Me}}$ < Cp$^{1\mbox{-}t\mathrm{Bu}}$. It therefore is
evident that ligand size alone is insufficient to provide a simple, unambiguous
relationship for the variation of $\omega$ in chemically related complexes. We
advise the reader to be aware that consideration of the angle between the two
cy\-clo\-pen\-ta\-di\-enyl rings and their torsion angle would be important for
a complete analysis.\cite{CR2018} But the exploration of such intricate details
requires a different data set of its own, one which is well beyond the scope of
this work. Here, we are focused on providing systematic extraction of
spin-crossover energies from available experimental {\thalf} values for a
chemically diverse set of metal complexes and the implications drawn therefrom
for various methodological options.

\begin{figure}
\centering
  \includegraphics[width=0.5\columnwidth]{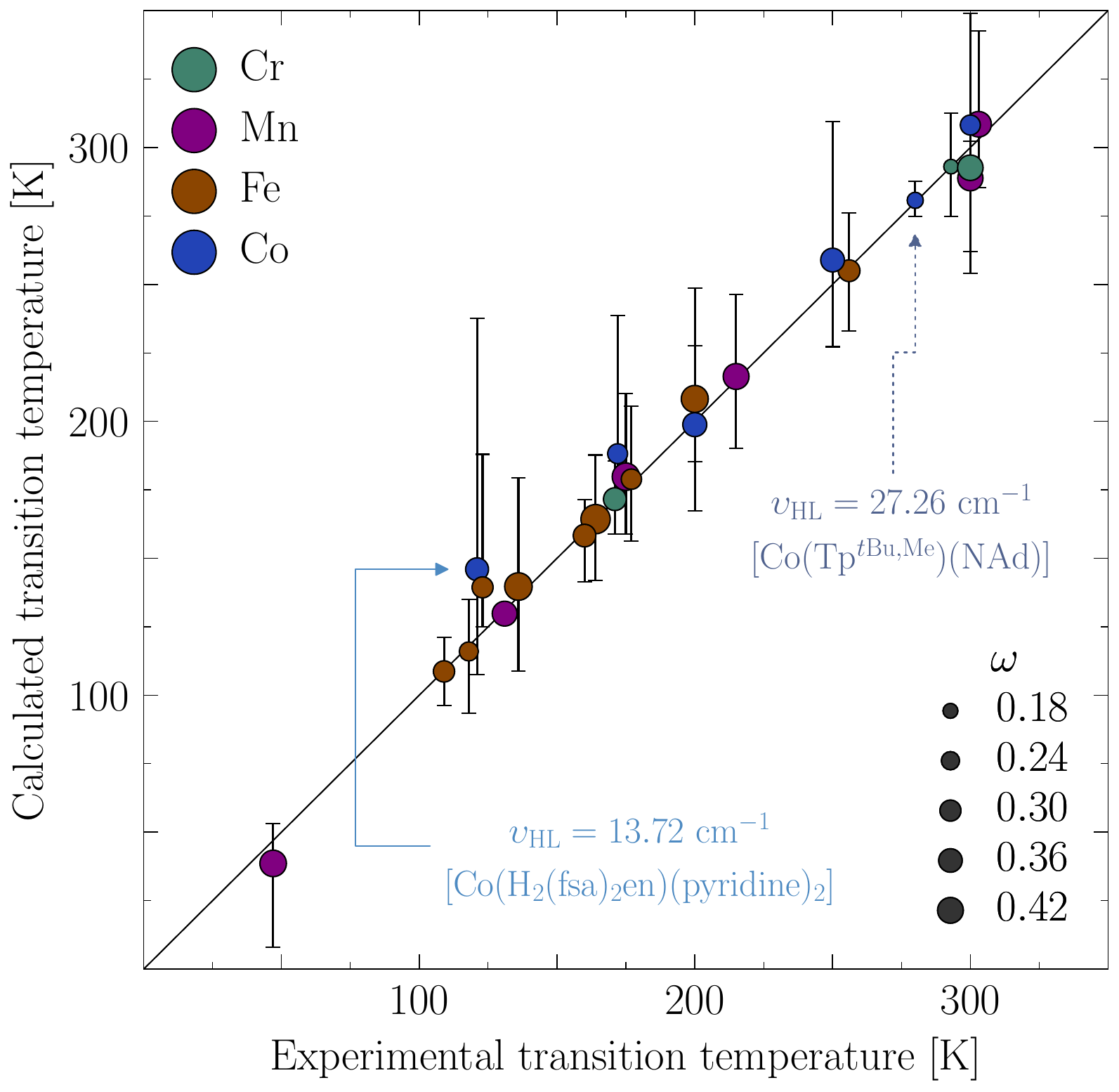}
  \caption{Comparison between the experimental and calculated transition
           temperature. For each species, the $\omega$ parameter in the
           LC-$\omega$PBE functional was optimized to reproduce as closely as
           possible the experimental target. The magnitude of these optimized
           parameters, as illustrated, is proportional to the size of the filled
           circle. The vertical bars, on the other hand, show the sensitivity of
           the calculated transition temperature to a $\pm$ 1 \% alteration of
           the $\omega$ parameter, whereas $\upsilon_\mathrm{HL}$ shows the
           harmonic mode with the lowest frequency mode at the equilibrium
           temperature for two representative examples. A softer harmonic mode
           results in a larger error bar as opposed to a harder mode.}
  \label{fig:1}
\end{figure}
  To illustrate the relative precision of the fitted $\omega$ values, Figure
\ref{fig:1} is a parity plot that compares the experimental and computed {\thalf}
values. The data themselves are included in Table \ref{tab:I} and in the
Electronic Supplementary Information. Observe that there is good correspondence
between the two, hence a successful $\omega$ parametrization that validates the
appropriateness of the reference electronic structure calculations.

  There remains, however, a valid procedural issue, specifically the
sensitivity of the calculated {\thalf} with respect to $\omega$. To assess that,
we perturbed the fitted $\omega$ by one percent both up and down for each metal
complex, and calculated the {\thalf} associated with those altered $\omega$
values. Those altered {\thalf} provide the error bars presented in Figure
\ref{fig:1}. It is striking, at least, that a seemingly rather small one percent
change in the range separation parameter $\omega$ leads to differences in
{\thalf} from as little as one Kelvin for {\scoC} to as large as 92 K for {\scoQ}.
Notice further that a larger sensitivity to alteration of $\omega$ corresponds
to a slightly reduced agreement between the optimally calculated {\thalf} and
its experimental datum.

  A deeper analysis of the origins of the {\thalf} sensitivity thus is
pertinent. We focus attention on the two complexes highlighted in Figure
\ref{fig:1}, namely, {\scoQ} and {\scoV} with calculated transition temperatures
of 146 K and 281 K, respectively. The reason behind this choice is their
distinct sensitivity to perturbation of $\omega$. In detail, the calculated
transition temperatures span the interval $108 \leq \thalf \leq 238$ K, or
$\Delta T = 130$ K, for the former complex versus a much smaller interval of
$275 \leq \thalf \leq 288$, or $\Delta T = 13$ K, for the latter one. A
distinguishing structural feature is that {\scoV} shows considerable steric
hindrance from the Tp$^{t\mathrm{Bu},\mathrm{Me}}$ and a\-da\-man\-tyl az\-ide
ligands, as opposed to the H$_2$(fsa)$_2$en and pyridine groups forming {\scoQ}.
An indirect measure of the comparative structural stiffness follows from the
smallest harmonic vibrational mode at the equilibrium temperature {\thalf}. For
{\scoQ}, there is a soft equilibrium mode of 13.72 cm$^{-1}$, but a harder mode
of 27.26 cm$^{-1}$ for {\scoV}. The difference indicates that the structural
stiffness for the former complex is relatively less prominent than that for the
latter and, in consequence, {\thalf} for {\scoQ} varies more rapidly with respect to 
changing $\omega$.

  Some of the other complexes that show notable sensitivity to perturbation of
$\omega$ are {\scoS}, {\scoR}, and {\scoW} with $\Delta T$ of 82, 69, and 64 K,
associated with a lowest equilibrium harmonic mode frequency of 11.62, 17.07,
and 14.63 cm$^{-1}$, respectively. Conversely, the metal complexes {\scoD},
{\scoA}, and {\scoN} show less sensitivity to $\omega$ perturbations, with a
$\Delta T$ of 2, 27, and 30 K, and lowest equilibrium mode frequencies of 19.75,
20.05, and 25.41 cm$^{-1}$, respectively. This collection of results constitutes 
evidence that the determination of {\thalf} via equation \eqref{eq:2} is more
susceptible to procedural changes for those complexes with relatively soft
equilibrium harmonic modes. There are noticeable exceptions nonetheless, for
instance, {\scoI} with $\Delta T = 71$ K and {\scoL} with $\Delta T = 25$ K.
The magnitudes of their smallest equilibrium harmonic frequencies are 8.01 and
4.76 cm$^{-1}$, respectively. Both are in clear contradiction with the
supposition about lowest harmonic frequencies. Both shed light upon the many
complicated nuances for accurate determination of transition temperatures.

  Given this difficulty of analysis, it is important to note that Table
\ref{tab:I} collects as well the computed $\Delta H$ and $\Delta S$ that relate
to {\thalf} through equation \eqref{eq:2}. Experimental availability for these
two quantities unfortunately often is limited  or non-existent. That
aside, before proceeding we need to ensure that our fitting to the experimental
equilibrium temperatures results in data that are thermodynamically meaningful.
For that, refs. \citenum{R2019, VFCR2020, DSR2021} proposed the use of
$\Delta H_\mathrm{expt} - \Delta H_\mathrm{calc}$ as the means for method
benchmarking. In that regard, ref \citenum{RMFS1997} reports $\Delta H = 13.4$
kJ\,mol$^{-1}$ for {\scoN}, whereas Table \ref{tab:I} shows $\Delta H = 14.49$
kJ\,mol$^{-1}$, or a deviation of +1.09 kJ\,mol$^{-1}$. Analogously for {\scoO},
ref \citenum{NGMABR2003} reports $\Delta H = 4.08$ kJ\,mol$^{-1}$, while our
fit differs by only +0.70 kJ\,mol$^{-1}$. In addition, the experimental data
for the {\scoR} complex from ref \citenum{GMNR2001} shows $\Delta H = 3.45$
kJ\,mol$^{-1}$, whereas the computed value in Table \ref{tab:I} differs only by
$-$0.1 kJ\,mol$^{-1}$. An analogous comparison for $\Delta S$ results in
deviations of +7.63, +7.18, and $-$2.18 J\,mol$^{-1}$\,K$^{-1}$ for {\scoN},
{\scoO}, and {\scoR}, respectively. These small differences show that the
parametrized $\omega$ used for computing {\thalf} and the associated electronic
structure calculations may be quite useful in detailed study of a given
candidate material, including calculation of the intensities for the
vibrational modes.

\begin{figure}
\centering
  \includegraphics[width=0.5\columnwidth]{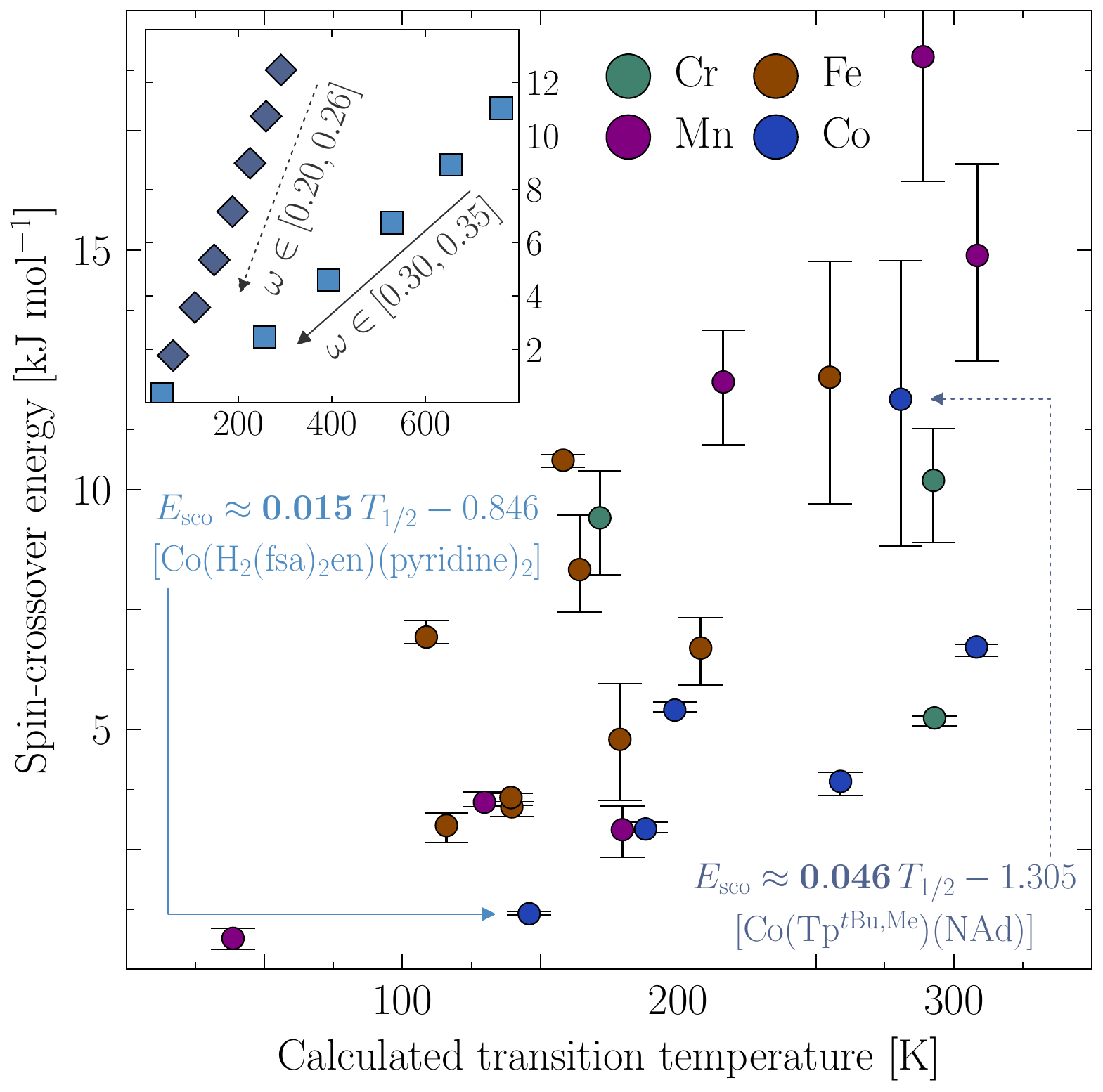}
  \caption{The spin-crossover energy, in kJ mol$^{-1}$, including the zero-point
           energy correction, computed with the optimized $\omega$ parameter in
           the LC-$\omega$PBE functional as a function of the calculated
           transition temperature, in Kelvin, for the set of metal complexes.
           The vertical bars depict the sensitivity of the spin-crossover energy
           to a $\pm$ 1 \% alteration in the transition temperature. The equations
           show a linear regression of ${\Delta E}_\mathrm{sco}$ vs $T_{1/2}$,
           depicted in the inset, using distinctly different ranges of $\omega$
           for two representative examples. A larger slope suggests a larger error
           bar and vice versa.
  \label{fig:2}}
\end{figure}
  Considering however the computational time and effort required for harmonic
frequency analyses for both the low- and high-spin states for every species of
interest in a high-throughput workflow, as well as the fact that many
functionals yield harmonic modes with rather inaccurate
intensities,\cite{AZZT2010} we return to an issue mentioned already,
namely the systematic extraction of a molecular quantity {\desco} from
experimental thermodynamic data, namely the measured {\thalf}. This reverse
engineering of {\desco} has a simple motivation, namely, that computation
of molecular energy differences is orders of magnitude faster than running
vibrational frequency analyses. Thus, high-throughput screening would benefit
greatly by reduction of the problem from one focused on {\thalf}, which can be
measured fairly routinely, to one focused on {\desco} for which there essentially
is no direct experimental access.

  The protocol by which we have tuned $\omega$ to yield a good fit to
experimental {\thalf} for each species includes all of the pieces necessary for
that reverse engineering. Figure \ref{fig:2} shows the resulting {\desco} values
as extracted from the calculated {\thalf}. We therefore refer to these {\desco} as
model-derived spin-crossover energies constrained by experimental temperatures.
The data also is available in the Electronic Supplementary Information,
and a Python code spinet to retrieve it is shown in Figure S1. The
individual results, including for the spin gap {\dehl}, are collected in Table
\ref{tab:I}. It is striking confirmation of the quantitative achievement of
our scheme that the spin-crossover energies are no larger than 20 kJ\,mol$^{-1}$.
That is an empirical threshold known in the literature.\cite{K2019} In fact, all
but one species have $\desco \leq 15$ kJ\,mol$^{-1}$.

  It clearly is of methodological importance to analyze the sensitivity of the
derived {\desco} with respect to the reference {\thalf}. To establish a
confidence interval,  once again we used reverse engineering. We both increased
and decreased the reference transition temperature by one percent for each system
and extracted the corresponding {\desco}. Those results are the error limits 
depicted in Figure \ref{fig:2}. Evidently, the spin-crossover energies for most
systems remain stable relative to small variations in the reference transition
temperature. However, some show deviations nearly as large as 6 kJ\,mol$^{-1}$.
The origins of those deviations appear to be rooted in the chemical
characteristics and functional groups of the ligands in the metal complex rather
than in simple structural features. Attempts to anticipate the sensitivity of
{\desco} from {\thalf} with regard to methodological choices for any given
complex seems, therefore, to be a complicated and rather opaque task.

  To illustrate the previous statement, consider once again the complexes
{\scoQ} and {\scoV}. Recall from Figure \ref{fig:1} that the calculation of
{\thalf} for the former is substantially more sensitive to procedural choices
than for the latter. Figure \ref{fig:2} shows that for extraction of {\dehl}
from {\thalf} the two systems exhibit converse behavior, namely, the error bar
for {\scoV} is larger than that for {\scoQ}. The origins of this behavior may be
understood by analysis of how {\thalf} and {\desco} change while varying
the $\omega$ parameter in the LC-$\omega$PBE functional. A simple linear
fit to {\desco} vs. {\thalf} for both metal complexes is shown in the inset of
Figure \ref{fig:2}. Those used a larger range of $\omega$ values that was
distinct for each system. The fit reveals that the variation in the
spin-crossover energy mostly is rooted in how rapidly {\desco} changes with
$\omega$. That, recall, at the same time changes {\thalf}. The slope of 0.046
for {\scoV} is three times larger than that of 0.015 for {\scoQ} and, as a
result, {\desco} changes more abruptly for the former complex. The variations
for {\desco}, {\thalf}, $\Delta H$, and $\Delta S$ for different choices of
the $\omega$ parameter are shown in Figures S2 and S3, and collected in Table S1
as well, in the Electronic Supplementary Information for all species in the data set.

  Prior knowledge of the specific sensitivity of both the spin-crossover energy
and transition temperature clearly is complicated, though it may offer an
opportunity for machine learning or deep learning models. In the mean time,
given that the average error in Figure \ref{fig:2} associated with the reverse
engineered {\desco} for our data set is 1.5 kJ\,mol$^{-1}$, it is reasonably
safe to expect an uncertainty of around $\pm 2$ kJ\,mol$^{-1}$ for
spin-crossover energies. That, as discussed for Figure \ref{fig:1}, pairs with
the uncertainty of roughly $\pm 50$ K for transition temperatures. Notice that
these error margins persist despite purposely fitting the calculated {\thalf} to
reproduce as closely as possible the experimental value. The message therefore
is clear: one should expect large uncertainties in common electronic structure
calculations meant for regular production.

\begin{figure}
\centering
  \includegraphics[width=0.5\columnwidth]{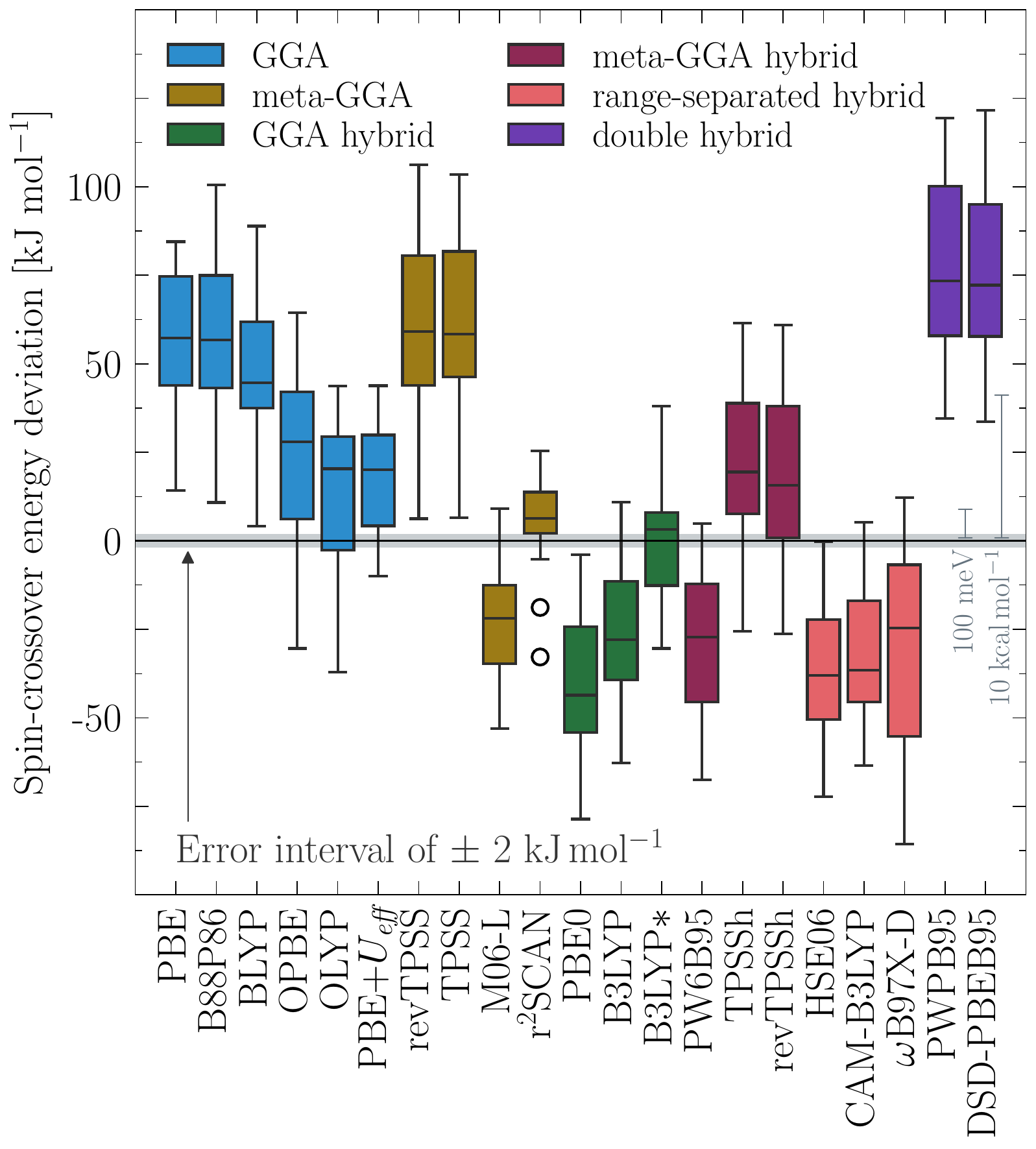}
  \caption{Deviation of the calculated spin-crossover energies using a
           selection of different choices of generalized gradient density
           functional approximations (GGAs), meta-GGAs, and various choices
           of hybrid functionals. The reference values are those computed with
           the optimized $\omega$ in LC-$\omega$PBE functional. For context, the
           boxes span the inter-quartile range, whereas the ends of the whiskers
           show the minimum and maximum deviation. The horizontal line in each
           case shows the median deviation for that functional, and the empty
           dots show the outliers.
  \label{fig:3}}
\end{figure}
  With that in mind, now we proceed to assess different generalized gradient
approximation density functionals (GGAs), meta-GGAs, global and
range-separated hybrids of them, and double hybrids.\cite{PS2001} Again keep in
mind that  all but one species in the reference set have $\desco \leq 15$
kJ\,mol$^{-1}$. Figure \ref{fig:3} presents the deviation of the spin-crossover
energy with respect to the reference data in Table \ref{tab:I} for a
selection of representative functionals. It is clear from that Figure
\ref{fig:3} that none of those functionals is capable of delivering an accuracy
of at least $\pm 15$ kJ\,mol$^{-1}$, or roughly a 100 \% relative error.

  Starting from the GGA level of refinement, Figure \ref{fig:3} shows that the
most widely used functional, PBE, results in a median deviation of 57.3
kJ\,mol$^{-1}$. The B88P86 approximation behaves alike, albeit with a slightly
wider distribution, whereas BLYP improves upon PBE by 12.7 kJ\,mol$^{-1}$. OPBE
and OLYP improve further with median deviations of 30.0 and 20.4 kJ\,mol$^{-1}$,
respectively, but with some unrealistic negative spin-crossover energies. The
partial improvement comes as no surprise since both of those functionals include
the OPTX\cite{HC2001} exchange density functional. It has been argued to
provide balanced accuracy for spin-crossover systems.\cite{SGEL2004, CG2007, S2008}
However, unlike most other constraint-based exchange functionals, OPTX does not
recover the correct homogeneous electron gas limit. Some of the limitations of
PBE may be overcome with the use of PBE+$U_\mathit{eff}$. In the form used here, there
is a Hubbard-$U$ correction for every species calculated by means of the linear
response for an ensemble average of the low- and high-spin
states.\cite{AHT2022} The resulting median deviation of 20.1 kJ\,mol$^{-1}$
is nearly the same as for OLYP. Note that PBE+$U_\mathit{eff}$ also may yield
negative energy differences, though experience thus far suggests that they are
not common. On a cost-benefit basis, we deem PBE+$U_\mathit{eff}$ as the best
performing among the generalized gradient approximations.

  Moving onward to the meta-GGAs, unsurprisingly revTPSS and TPSS nearly are
identical in behavior, each with a median deviation of 59.0 and 58.4
kJ\,mol$^{-1}$, respectively. The difference between the two is that, relative
to TPSS, revTPSS restores the second-order density gradient expansion for
exchange and uses a fitted density-dependent correlation gradient
coefficient.\cite{HL1986} Though such additional constraint satisfaction
generally is beneficial, in this instance it is ineffectual.\footnote{Both TPSS
and revTPSS have an order of limits problem.\cite{RSXC2012} Whether it is
relevant we do not know.} The empirical M06-L mostly underestimates the
spin-crossover energies, with a median deviation of $-$21.9 kJ\,mol$^{-1}$.
Note, however, that several works found this meta-GGA somewhat useful for spin
gaps.\cite{ZT2020, RDHS2024} Then there is r$^2$SCAN, currently a prominent
non-empirical, constraint-based meta-GGA functional. It provides a median
deviation of 6.25 kJ\,mol$^{-1}$. Nonetheless it is worrisome that for
two species r$^2$SCAN over-stabilizes the high-spin state relative to its
low-spin partner. That generates large magnitude, wholly unrealistic negative
energy differences, as clearly seen with the two outliers registered as empty
dots in Figure \ref{fig:3}.

  In regards to the global GGA hybrids PBE0, B3LYP, and B3LYP$\ast$, global
meta-GGA hybrids PW6B95, TPSSh, and revTPSSh, and range-separated hybrids HSE06,
CAM-B3LYP, and $\omega$B97X-D, it is clear from Figure \ref{fig:3} that the
majority of them give non-physical negative spin-crossover energies, with PBE0
giving the largest magnitude median deviation of -43.6 kJ\,mol$^{-1}$. The two
notable exceptions within this group are TPSSh and revTPSSh, with median deviations
of 19.4 and 15.7 kJ\,mol$^{-1}$, respectively. Their main structural characteristic
is that both include a 10 \% fraction of single determinant exchange, as distinct
from the 25 \% contribution in PBE0 or 20 \% in B3LYP. The 15 \% contribution
in B3LYP$\ast$ reduces the error relative to B3LYP, with a median of 3.3
kJ\,mol$^{-1}$, but still results in unrealistic negative spin-crossover energies
for several species. The results for all of these global hybrids demonstrate
clearly that the larger the fraction of single determinant exchange, the more
negative the {\desco}. This relationship is one of the motivations for use of
B3LYP$\ast$, TPSSh, and revTPSSh for these complexes.\cite{SYMBY2014,
CVR2018, CR2020}

  The last two double hybrids PWPB95 and DSD-PBEB95, in Figure \ref{fig:3}, are
the worst performing of all the functionals we considered,
with a median deviation of 73.4 and 72.2 kJ\,mol$^{-1}$, respectively. Those large
deviations are unsurprising because these two double hybrids are heavily
parametrized for use in thermochemistry and kinetics with the GMTKN30
database.\cite{GG2011a} It mostly is comprised of small organic and inorganic
molecules. We return to that issue in the Conclusions.

\begin{figure}
\centering
  \includegraphics[width=0.5\columnwidth]{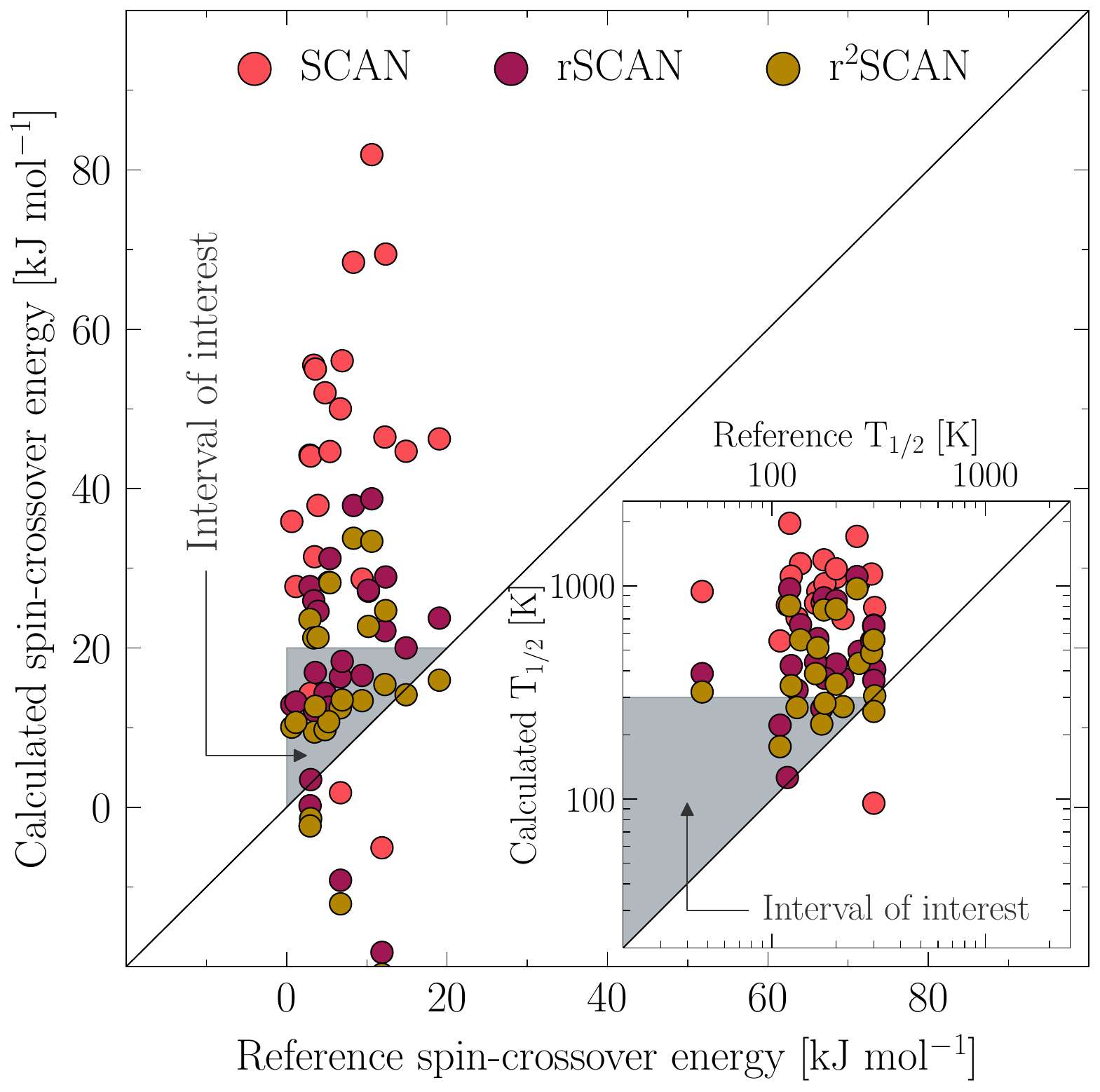}
  \caption{Correspondence between the calculated spin crossover energies for the
           SCAN, rSCAN, and r$^2$SCAN meta-generalized gradient
           approximations with respect to the reference values using the
           LC-$\omega$PBE density functional with the $\omega$ parameter optimized
           for each metal complex. The inset shows the analogous comparison, but
           for the transition temperature, $T_{1/2}$, instead.
  \label{fig:4}}
\end{figure}
  From the perspective of functional development based on constraint satisfaction,
and in consideration of the modest accuracy observed for the r$^2$SCAN meta-GGA,
exploration of the triad SCAN, rSCAN, and r$^2$SCAN is a reasonable endeavor.
Figure \ref{fig:4} compares {\desco} for the three with respect to our reference
values in Table \ref{tab:I}. The error associated with each functional follows
the trend r$^2$SCAN < rSCAN < SCAN. Notice further that many spin-crossover
energies are either too large and a few are negative. In the case of overly
large calculated {\desco}, the inevitable consequence is an excessive calculated
{\thalf}. That is shown in the inset for Figure \ref{fig:4}. In multiple cases,
the transition temperatures from all three meta-GGAs are several hundred Kelvin
too high with respect to the reference values in Table \ref{tab:I}. Second, if
the calculated $\desco < 0$, we cannot even associate a calculated transition
temperature for that metal complex. There is no equilibrium {\thalf} in going
from a low-spin ground state to a high-spin state, since the ground state for a
$\desco < 0$ already is high spin.\footnote{There are examples in the literature
for such reverse spin conversions, but descriptions for those mostly require the
use of phenomenological models.\cite{BSHV2000}}

  As it stands, even sophisticated state-of-the-art approximate functionals, like
r$^2$SCAN, lack the predictive precision required for realistic thermodynamic
description of thermally driven spin-crossover complexes. The main reason is
that the range $0 \leq \desco \leq 15$ kJ\,mol$^{-1}$ is substantially out of
reach. That is true even with what Perdew and Schmidt call the heaven of chemical
accuracy, for which errors in computations approximately are of 4 kJ/mol.\cite{PS2001}
Such magnitude already accounts for a 27 \% relative error with respect to the
target {\desco}, an error that, as shown, results in unrealistic transition
temperatures. Put differently, the underlying challenge is that the relevant
spin-crossover energetics are comparable in magnitude to thermal energy scales
near room temperature. As a result, small kJ\,mol$^{-1}$ errors in computed
energy differences can lead to substantial deviations in predicted transition 
temperatures.

\begin{table}
  \centering
  \caption{Number of basis functions, $N_b$, number of electrons, $n_e$, and
           total energies differences $E_\mathrm{CCSD(T)} - E_\mathrm{CCSD}$,
           in kJ\,mol$^{-1}$, computed with the coupled cluster singles and
           doubles method, CCSD, and with the addition of perturbative triples,
           CCSD(T), for the set of eight metal complexes. The resulting spin gap
           energy {\dehl} in units of kJ\,mol$^{-1}$ for the baseline
           Hartree-Fock, HF, and for both methods also is included for
           comparison. All geometries were relaxed with the meta-generalized
           r$^2$SCAN density functional approximation. \label{tab:II}}
  \begin{adjustbox}{max width=\columnwidth}  
  \begin{threeparttable}
  \begin{tabular}{ll
                  S[table-format=4.0]
                  S[table-format=3.0]l
                  S[table-format=-3.2]
                  S[table-format=-3.2]l
                  S[table-format=-3.2]
                  S[table-format=-3.2]
                  S[table-format=-3.2]}
    \hline
    {Complex}&&{$N_b$}&{$n_e$}&&\multicolumn{2}{c}{$E_\mathrm{CCSD(T)} - E_\mathrm{CCSD}$}
                              &&\multicolumn{3}{c}{\dehl}\\ \cline{1-1}\cline{3-4}\cline{6-7}\cline{9-11}
             &&       &       &&{Low spin}&{High spin}&&{HF}&{CCSD}&{CCSD(T)}\\ 
    {\extD}  && 1336  & 228   &&-791.35&-761.10&&-193.76& -30.19&   0.05\\
    {\scoE}  &&  660  & 111   &&-444.92&-340.65&&-624.44&-116.11& -11.84\\
    {\scoF}  && 1008  & 159   &&-582.85&-415.14&&-633.03&-115.85&  51.86\\
    {\extH}  && 1156  & 193   &&-642.90&-585.32&&-358.47& -66.60&  -9.03\\
    {\extT}  &&  452  & 83    &&-212.20&-136.72&&-391.31&-193.52&-118.04\\
    {\extU}  &&  800  & 125   &&-371.85&-319.32&&-244.39&  -0.30&  52.23\\
    {\extM}  && 1316  & 244   &&-954.46&-770.22&&-376.55& -55.08& 129.16\\
    {\extV}  && 1064  & 168   &&-505.23&-469.66&&-110.43&  66.69& 102.27\\
    \hline 
    \end{tabular}
  \begin{tablenotes}
  \setlength\labelsep{0pt}
  \item\scriptsize{L$_2$ = (2,2$^\prime$\--(2,6,9,13\--te\-tra\-aza\-te\-tra\-de\-ca\--1,13\--di\-ene\--1,14\--di\-yl)\-di\-phe\-nol),
    Cp$^{1\mbox{-}\mathrm{Me}}$ = me\-thyl\-cy\-clo\-pen\-ta\-di\-ene,
    Cp$^{1\mbox{-}t\mathrm{Bu}}$ = \emph{tert}\--bu\-tyl\-cy\-clo\-pen\-ta\-di\-enyl,
    acac = a\-ce\-tyl\-a\-ce\-to\-na\-te, tr\-ien = tri\-e\-thyl\-ene\-te\-tra\-mi\-ne,
    eda = eth\-yl\-ene\-di\-a\-mi\-ne,
    bpp = (2,6\--di(py\-ra\-zol\--1\--yl)\-py\-ri\-di\-ne),
    tacn = 1,4,7-tri\-a\-za\-cy\-clo\-non\-ane}
  \end{tablenotes}
  \end{threeparttable}
  \end{adjustbox}
\end{table}
  Given this rather unsatisfactory situation with Kohn-Sham methodology and
contemporary functionals, we turn to the possibility of using the coupled cluster
method in its typical form, CCSD(T), mentioned already. Though computationally
costly and beset with formidable cost scaling, in both concept and computational
implementation CCSD(T) is completely distinct and separate from the challenges
intrinsic to the design and implementation of density functionals. As a practical
matter of cost, geometry optimization of any of the metal complexes
considered here is beyond reach. Thus, it was reasonable to exploit the best of
both Kohn-Sham and coupled cluster. We started with the r$^2$SCAN meta-GGA for
structural relaxation. For those geometries, the reference unrestricted Hartree-Fock
calculation was done, with the resulting self-consistent Hartree-Fock orbitals used
for the eventual CCSD and CCSD(T) calculations.

  The results for this computational procedure are shown in Table \ref{tab:II}
for the eight-system set of metal complexes. The total numbers of basis functions
and electrons also are included for the sake of reference. Given that these series
of calculations used a different choice of basis set, small variations to
{\dehl} are inevitable, but expected within the uncertainty range of $\pm 2$
kJ\,mol$^{-1}$. The first thing to notice is that the spin gaps computed
with Hartree-Fock go from $-$633.03 to $-$110.43 kJ\,mol$^{-1}$, thus are grossly
underestimated. This is not unexpected in light of the strong tendency of
Hartree-Fock to over-stabilize high-spin states.\cite{PFHP2018} What this
illuminates is the formidable barrier to good {\dehl} values that is intrinsic
to conventional CCSD or CCSD(T) calculations by the normative use of a
Hartree-Fock reference.

  Consistent with that insight, the spin gap energies computed with CCSD show a
sharp improvement over the underestimations given by Hartree-Fock. The CCSD
results span the interval $-$193.52 to 66.69 kJ\,mol$^{-1}$. Though improved, the
range still remains far from the reverse-engineered reference data in Table
\ref{tab:I}. The problem is qualitative. All complexes but {\extV} show a
negative spin gap, though the {\extU} spin gap energy of $-$0.30 kJ\,mol$^{-1}$
is on the verge of being positive. Put stringently, at best six of the eight
have the wrong sign for {\dehl} from CCSD.  Further inclusion of perturbative
triple excitations, CCSD(T), fixes the sign for four complexes, namely, {\extD},
{\scoF}, {\extU}, and {\extM}. Though still improperly negative, the spin gap
energies of $-$11.84 and $-$9.03 kJ\,mol$^{-1}$ for the {\scoE} and {\extH}
complexes, respectively, also benefited significantly in going from CCSD to
CCSD(T). The absolute magnitude of {\dehl} for the remaining three systems
persists at larger than 100 kJ\,mol$^{-1}$.

  Judging by the tendencies exhibited in Table \ref{tab:II}, there are several
points of criticism regarding the coupled cluster calculations. As
expected,\cite{LP2007, HTLWH2019, DMP2022, FP2022} the total energy difference
$E_\mathrm{CCSD(T)} - E_\mathrm{CCSD}$ for the series CCSD, CCSD(T) is not
converged with respect to excitation order, apparently primarily due to the
inadequateness of the unrestricted Hartree-Fock reference and the possible
multi-reference nature of some of the spin states of these metal complexes.
Because the molecular spin gap energy
of interest is the difference of two comparatively large magnitude total
energies, it is relevant that the conventional coupled cluster implementation we
used is not variational, which may worsen such differences. This sort of problem
is not uncommon. We emphasize that our calculations are large compared to more
common coupled cluster calculations and that achieving energy convergence even
for these medium-sized spin-crossover metal complexes was nearly prohibitive.
The challenges are both the intensive demand of computational resources and the
long run times even on exascale machines. Extrapolation to the complete basis set
limit therefore also is out of reach for now.

%%%%%%%%%%%%%%%%%%%%%%%%%%%%%%%%%%%%%%%%%%%%%%%%%%%%%%%%%%%%%%%%%%%%%%%%%%%%%%%%
\section{Conclusions \label{sec:conclusions}}
%%%%%%%%%%%%%%%%%%%%%%%%%%%%%%%%%%%%%%%%%%%%%%%%%%%%%%%%%%%%%%%%%%%%%%%%%%%%%%%% 

  To summarize, we reproduced as closely as possible the transition temperature
for a set of spin crossover metal complexes by means of fitting the range
separation parameter $\omega$ in the LC-$\omega$PBE hybrid density functional.
Those calculations then were used for reverse engineering the associated
molecular spin crossover energy. Further sensitivity analyses
showed an uncertainty of $\pm 50$ K for transition temperatures, and $\pm 2$
kJ\,mol$^{-1}$ for spin crossover energies.

  Based on our comparative testing relative to the extracted
{\dehl} data set, the accuracy of several
exchange and correlation functional approximations appears to be
insufficient for predicting {\dehl}, not even achieving modest
deviations of $\pm 15$ kJ\,mol$^{-1}$ with respect to our reference
values. That already accounts for at least a 100~\% relative error. As
a result, such limited accuracy may impede a realistic electronic and
thermodynamic description of the thermally driven spin crossover
phenomenon. At present the best available cost-benefit compromise
seems to be to study trends within chemically similar sets of
complexes using either PBE+$U_{\mathit{eff}}$ or r$^2$SCAN.

  In view of these limitations of non-empirical constraint-based density
functionals, the question arises as to whether the spin crossover phenomenon
should be studied with empirical or substantially parametrized functionals. We
have serious reservations about that because of the near-impossibility of
detecting and controlling unintended bias introduced by the parametrization.
That illustrates the need for and the eventual benefit from use of data sets
with transition metals for benchmarking purposes in the development of
functionals. Such data sets are, to our knowledge, regrettably scarce, with
only a few available in the literature.\cite{IJ2019, ZT2020, BS2020} For
calibration, note the rather marginal number of species with transition metals
in the newest GSCDB137 data set.\cite{LH2025}

  The more computationally intensive coupled cluster calculations for a set of
eight metal complexes proved challenging. Our results seem to indicate a
non-systematic convergence for the truncation of the excitation expansion, with
inclusion of perturbative triples appearing to show signs of early convergence.
We therefore suggest additional in-depth studies for a larger data set in
regards to the sensitivity of the choice of basis set, larger excitation
expansions, and the use of different mean-field references.

  Overall, the reader should be acutely aware that there will be downsides to
any electronic structure calculation of spin crossover. The affordability of
Kohn-Sham calculations is compromised by the uncertainties introduced by the
design choices of whatever functional approximation is selected. Our calculations
show that such uncertainties can be quite large for metal complexes. While the
converse is true in principle for high-level wave function methodology such as
coupled cluster schemes, in practice the matters of choices of reference function,
excitation-level convergence, and enormous demands on computational resources
makes realization of that in-principle superiority virtually unachievable in
practice. Nevertheless, we hope that our reverse-engineered molecular spin
crossover energies will be a useful benchmark to assess the reliability of
different high-level wave function methods.

  Lastly, we hope that, in light of our results, future development of 
exchange and correlation approximations considers more seriously the use of test
sets with realistic transition metal systems.

%%%%%%%%%%%%%%%%%%%%%%%%%%%%%%%%%%%%%%%%%%%%%%%%%%%%%%%%%%%%%%%%%%%%%%%%%%%%%%%%
\section*{Data and Software Availability \label{sec:data}}
%%%%%%%%%%%%%%%%%%%%%%%%%%%%%%%%%%%%%%%%%%%%%%%%%%%%%%%%%%%%%%%%%%%%%%%%%%%%%%%%

  The public repository \href{https://doi.org/10.5281/zenodo.20526833}{doi.org/10.5281/zenodo.20526833}
collects all the information computed for the twenty-four neutral metal complexes
in the data set, for both the low- and high-spin states, using the optimized
$\omega$ parameter for the range separated \mbox{LC-$\omega$PBE} density functional.
These entries include the Cartesian coordinates in {\AA} for each species; the
calculated spin crossover energy ${\Delta E}_\mathrm{sco}$ in kJ\,mol$^{-1}$; the
change in enthalpy $\Delta H$, and change in entropy $\Delta S$, in units of
kJ\,mol$^{-1}$ and J\,mol$^{-1}$\,K$^{-1}$, respectively; and the calculated
transition temperatures in Kelvin. Additional data such as the total energy in eV,
harmonic frequencies in Hz, orbital energies in eV, various other arrays with
local spin, atoms, elements, as well as the rotational symmetry, and rotational
temperature also are provided to reproduce the reported properties, as shown in
Figure S1 in the Supplementary Material, using the open source pySCO library
available in the public repository
\href{https://github.com/amalbavera/pysco}{github.com/amalbavera/pysco}.

%%%%%%%%%%%%%%%%%%%%%%%%%%%%%%%%%%%%%%%%%%%%%%%%%%%%%%%%%%%%%%%%%%%%%%%%%%%%%%%%
\section*{Supplementary Material \label{sec:esi}}
%%%%%%%%%%%%%%%%%%%%%%%%%%%%%%%%%%%%%%%%%%%%%%%%%%%%%%%%%%%%%%%%%%%%%%%%%%%%%%%%

  The Electronic Supporting Information depicts a code snippet using Python to
retrieve the information available in the public repository
\href{https://doi.org/10.5281/zenodo.20526833}{doi.org/10.5281/zenodo.20526833}
to reproduce the data shown in Table \ref{tab:I}, and collects the
spin-crossover energy, transition temperature, change in enthalpy, and change
in entropy as function of the parameter $\omega$ in the LC-$\omega$PBE
functional for all the metal complexes in the data set.

\begin{acknowledgement}
  The authors are grateful to Ajith Perera for helpful discussions and his
insights into the coupled cluster calculations. We also extend our thanks to
Edoardo Apr{\`a} for his continuous support and invaluable advice that helped
us run the calculations in the exascale facilities.

  AAM, RGH, and SBT were supported as part of the Center for Molecular Magnetic
Quantum Materials, an Energy Frontier Research Center funded by the U.S.
Department of Energy, Office of Science, Basic Energy Sciences under Award No.
\mbox{DE-SC0019330}. DMR and NG acknowledge support from the U.S. Department of
Energy, Office of Science, Basic Energy Sciences, Division of Materials Science
and Engineering, Theoretical Condensed Matter Physics Program under
\mbox{FWP 83557}. AP acknowledges support from the U.S Department of Energy,
Office of Science, Office of Basic Energy Sciences, division of Chemical Sciences,
Geosciences and Biosciences under \mbox{FWP 79715} at the Pacific Northwest
National Laboratory (PNNL). This research used resources of the National Energy
Research Scientific Computing Center (NERSC), a Department of Energy Office of
Science User Facility using NERSC award \mbox{BES-ERCAP0022828}, and resources
of the Oak Ridge Leadership Computing Facility Frontier and Argonne Leadership
Computing Facility through the ASCR Leadership Computing Challenge (ALCC)
project \mbox{ALCC [2024-2025]} Exploring Exascale Quantum Chemical Methods for
Transition Metal Chemistry.
\end{acknowledgement}

%%%%%%%%%%%%%%%%%%%%%%%%%%%%%%%%%%%%%%%%%%%%%%%%%%%%%%%%%%%%%%%%%%%%%%%%%%%%%%%%
\bibliography{references}
%%%%%%%%%%%%%%%%%%%%%%%%%%%%%%%%%%%%%%%%%%%%%%%%%%%%%%%%%%%%%%%%%%%%%%%%%%%%%%%%

\end{document}